\documentclass[aps,pre,floatfix,twocolumn,nofootinbib,showpacs]{revtex4} 
\usepackage{graphicx}\usepackage{amsmath}\usepackage{amssymb}

\newcommand{\Nw}{N_{w,oe}^{e}} 
\newcommand{\Nc}{N_{cy,oe}^{e}}
\newcommand{\Mw}{M_{w,ot}^{e}}

\newcommand{\Mwt}{M_{w,ot}^{t}}

\newcommand{\Nctt}{N_{cy,ot}^{t}}

\newcommand{\Mwolet}{M_{w,o'e}^{t}}

\renewcommand{\tt}{t_t}
\newcommand{\te}{t_e}
\newcommand{\st}{s_t}
\newcommand{\se}{s_e}

\newcommand{\roee}{r_{oe}^{e}}

\newcommand{\cote}{c_{ot}^{e}}

\newcommand{\coeoee}{c_{oe,oe}^{e}}
\newcommand{\coeote}{c_{oe,ot}^{e}}
\newcommand{\cotote}{c_{ot,ot}^{e}}
\newcommand{\cotoee}{c_{ot,oe}^{e}}

\newcommand{\coeotr}{c_{oe,ot}^{r}}

\newcommand{\ct}{c_t}
\newcommand{\ce}{c_e}

\newcommand{\betae}{\beta_e}
\newcommand{\betat}{\beta_t}

\newcommand{\fcyoee}{f_{cy,oe}^{e}}
\newcommand{\fwoee}{f_{w,oe}^{e}}

\newcommand{\fwote}{f_{w,ot}^{e}}
\newcommand{\fwott}{f_{w,ot}^{t}}

\newcommand{\vt}{v_t}
\newcommand{\ve}{v_e}

\newcommand{\obst}{observer-train}
\newcommand{\obse}{observer-emb}
\newcommand{\Obst}{Observer-train}
\newcommand{\Obse}{Observer-emb}
\newcommand{\Oe}{O_{emb}}
\newcommand{\Ot}{O_{train}}

\newcommand{\BEQ}{\begin{eqnarray}}
\newcommand{\EEQ}{\end{eqnarray}}
\newcommand{\BEA}{\begin{eqnarray}}
\newcommand{\EEA}{\end{eqnarray}}
\newcommand{\nn}{\nonumber}

\begin{document} 
\title{Foundations of Special Relativity and the Principle of Conservation of Information}

\author{C. Pombo and Th. M. Nieuwenhuizen}

\address{Institute for Theoretical Physics,
University of Amsterdam, Valckenierstraat 65, 1018 XE Amsterdam, The Netherlands}

\begin{abstract}
The theory of special relativity can be generalized by means of a new principle called 
Conservation of Information. 
This allows a derivation of the constancy of the velocity of light with respect to moving frames,  and, 
consequently, of Einstein's special relativity. The analysis is based on a review of the concept of observer. 
It is put forward that observers are not uniquely defined and that an observational
asymmetry, defined by the different ways in which light influences observations, 
lies at the origin of the non-absolutism of time. 
This observational difference is a kinematic condition, not an exclusive result for light, implying 
that non-absolutism of time may have a cause different from the electromagnetic nature of light.
The Lorentz transformations are  derived and different concepts of the velocity of light, 
relative to different classes of observers, are considered. 
\end{abstract}


\maketitle
\section{Introduction}

The theory of special relativity is based mainly on three principles. The first is the
statement of the rectilinear uniform movement. The second is the constancy of the
velocity of light as it was named by Einstein. The third is a generalization of 
Newtonian dynamics. The two first laws are pure kinematic in their nature.
Alone, they can be represented by means of the two well known 
fundamental Lorentz transformations 
for time and space \footnote{The question why these laws can be expressed by 
means of transformations can be easily explained. 
Laws are statements for classes of phenomena, not for individual events. Then, considered as
a set of measures, a relation connecting the sets can 
be interpreted as a transformation. But it can also be interpreted as the 
description of a specific event, with respect to an arbitrary system of reference. In this way,
a transformation also represents the arbitrarily or generality of the event. The first relation
involving times, comes from the assumption of the constancy of the velocity
of the light in the expression of the relative velocity, as it is usually done.
From $c^2t^2-v^2t^2=w^2t^2$, setting $wt=ct'$, we get the first transformation. The
second one is just obtained by multiplying both sides by $v$ and substituting $s=vt$ and 
$s'=-vt'$. These correspondences can be found
in many text books and articles. See references ~\cite{pa1981,bo1962,re1968}.}.
\BEQ t=\frac{t'}{\sqrt{1-v^2/c^2}} \nn\EEQ 
and 
\BEQ s=\frac{s'}{\sqrt{1-v^2/c^2}}.\nn\EEQ

If understood correctly, these 
expressions alone already have 
the answer for a question which has been occupying the minds
of many physicists and philosophers since Einstein. This is the question about the origin 
of the constancy of the velocity of light, whether it is due 
to its electromagnetic nature
or not. We will explain, in simple terms, why the answer is certainly not. 

In pure classical physics, 
these transformations should be replaced by the Galilean 
transformations, which do not involve the velocity of the light. 
If we assume that $t'$ and $s'$ are sequences of measures of
proper time and proper length, which are just names for 
the classical measures, we are also assuming that these 
measures do not depend on light, but on
classical relations and standard units of measurements. Then
the equations state that the other two sets
of measures, $s$ and $t$, depend on the velocity of the light. 
Consequently, there are different classes of measures, the ones
depending on the velocity of the light
and the others which do not depend,
implying different ways of observing. 
And, if this is the case, there 
emerges the question of why certain observations depend on the velocity
of the light and others do not.

Before to focus on this question, two aspects of these relations 
deserve careful attention. Firstly, they only involve the velocity of what,
in separated words, is said 
to be the velocity of the light. Because velocity is a pure 
kinematic concept, in that concept of
velocity $c$ 
there is no indication about what moves. 
In general, velocity of  
wave is defined by means of wavelength and time, as 
$v_w=n(t)\lambda/t$,
where $n(t)$ is the number of waves considered
in an interval $t$. 
Since in this expression
there is no signal of electromagnetism,
we can assume that, in those 
expressions of Lorentz transformations, 
the velocity $c$ is not necessarily the velocity
of an electromagnetic wave.  

Measures of spatial distance are based on systems of reference
and Galilean relativity teaches us that
expressions of trajectories are relative to the velocity 
of the reference systems.
This does not involve change of unit or transformation of 
coordinates, which can be fixed with respect 
to all systems of reference.  Comparison
between measures also presupposes 
a common units. 
But when distances become dependent on an extra parameter such as
a velocity, there can be a change of unit, specified by
the movement of a certain entity. 

If this is not the case, the unit is fixed, then the
new parameter represents a new object of reference 
and there emerges the question of the uniqueness
of the event, when described by different observers. This is the second 
aspect to notice,
brought by the transformations above. If one observer describes 
an event including a certain entity, while the other
point of view does not, in which sense can we say that the
different points of view are about a unique event? In this case,
the observations
are qualitatively different, so we must speak about classes of
observers and we cannot define uniqueness, unless we define a new class of
observers, integrating both classes already present, unifying them. 
In this case, those transformations simply say that the uniqueness of events, with
respect to different classes of systems of reference, or observers,
does not occur at the same time. Considering a specific event, 
observations not depending on the travel of the light happen during 
a certain interval of time, while observations
depending on light happen during another interval of time. The different classes
of observation happen with different time durations.

These two aspects of the transformations only point to the optical 
nature of light,
independent of its electromagnetic nature. As an optical entity, light
can be a means for observations of distant events. In principle, this
does not depend on dynamical aspects of the
events and, consequently, of their electromagnetic features.
Originally, the rate $v/c$, which rules the fundamental conversion
of measures, is only a signal of conversion, an observational term,
not representing any term of interaction
between the systems of reference involved. It
does not take in account the
physical nature of what is being transfered between the observers. 
In conclusion, we do not have any {\it a priori} reason to assume
that the fundamental Lorentz transformations of space and time
have an electromagnetic origin. This is the reason why Einstein never derived
his special relativity from anything else but postulated it. The choice
of the formulation, if by means of the two laws or by means of two
fundamental Lorentz transformations, is irrelevant for us since they
are equivalent to each other. 
The fact that special relativity and Maxwell's electromagnetism
are both based on the non-absolutism of time
only shows that Maxwell's theory is a relativistic theory. 

In order to derive special relativity from another basis, Einstein had to explain
the origin of the non-absolutism of time, independent of
the constancy of the velocity of the light itself. 
This was his intention, when he discussed the relativism of the
simultaneity in his book `Relativity'~\cite{einsteinrel}. There,
Einstein introduces the 
question of simultaneity by means 
of an experiment which became known in the literature as the `train/embankment experiment'. 
The  experiment involves two 
beings named observers, one localized in the train and the other on the embankment. 
The train moves and these 
beings are reached by light coming from two separated but simultaneous 
lightnings striking the train. 
Einstein assumes that he explains the relativity of the simultaneity 
by means of analyzing the way in which light reaches the bodies of the
physical persons he considers as observers. However, a careful analysis
of his experiment, shows that several of Einstein's remarks 
about what these beings observe, are impossible to occur to physical observers in the 
conditions he describes. 

The main problem in Einstein's discussion was the hypothetical localization
of the observers, as argument for what they observe. 
This idea of localization generates serious misunderstandings 
because it suggests to the readers that our ways of thinking in physics can 
be a result of the places in which our bodies are settled. 
After Einstein, the vast literature concerning special relativity adopted the 
localization of observers as a didactic method\footnote{ There is a 
vast literature concerning
special relativity. We cite a few standard textbooks, in the end of the section of references.}. 
The localization of an observer as a physical person is supposed to 
explain why a local measure can be known by a specific person and not by
a second one supposed to be far from the first. In this explanatory context,
a measure can be known from a register without the need of transmission of light,
because the person, who is the one who knows, is so near the register that
can see, touch or hear it without delay in time or any kind of distortion due to 
relative movement. Another person cannot perceive the same signal, 
at the same time, because depends on reception of light for this.
The problem in this argument is that to perceive is not the same as to know.
To know, which in the context of physics is the same as to observe, is more than 
to measure or count. To know accounts for what is done or thought about the measures.
A pure measure is always an arithmetic element, without a physical
context and, consequently, it is not a physical observation. 
One single set of measures can give rise to more than one physical concept and 
this is the main reason why the localization of measures
does not give rise to specific observations. Two physical persons,
in different places, can think differently
about the same set of measures but a single person can interpret 
the same numbers in different ways. Consequently, it is not
localization of persons what produces different observations. 
In this way, a single person can play as two different observes and, consequently,
an observer is not a physical person. So, localization
of persons is far from explaining the fundamental point
of relativity. As far as we know, no analysis was carried out to elucidate 
the question of the non-absolutism of time  
as an observational fact. In the literature, special relativity
starts from the non-absolutism as a postulate, otherwise the constancy 
of the velocity of the light is the starting point.

Localization is not a condition of physical observers, but it can be a condition
of other kinds of observation. Physical observations are physical statements 
and depend on 
systems of reference, not on interaction between bodies of
persons and events, as it is the case for perceptions.
The latter are observations of the senses, not of physical
events outside. In both cases of observation, physical and perceptual,
reception of light can be a necessary element and these observers
belong to the class of the receivers of light. But, because reception of light
takes for granted light which is locally registered, receivers of light cannot
describe light waves or light rays which are in the space around
the register. This is the reason why this class of observers
never describes or observes light in space, but only far events which emit or 
reflect light. To register and to observe are not the same process,
observations can happen by means of light or without light and,
while reception of light is local, observation is not. This is why
localization of observers does not play role on observations.

In the train-embankment experiment, Einstein did not 
discriminate between classes of 
observers, because considering that something more than registration was
necessary for observations, it could generate doubt about the 
nature of the observers, if they actually could be replaced 
by registers and machines. However, 
registration alone does
not give rise to associations between local measures and
nonlocal ones, such as trajectories that are associations
between the sequence of local cycles of a clock and far distances. 
Einstein interpreted local reception of light from
distant sources as
the knowledge of distant source, without realizing that this
only can happen for receivers of light. In this way he assumed 
a direct correspondence between 
non simultaneity of local reception of light from two distant sources 
and non simultaneity of the separated emissions at the sources. But this does not happen
if one can observe the travel of the light. In his explanations, he went further
by considering vision as element of physical observation, stating that
his observers could see rays of light. This assumption is a serious mistake,
frequently found in the literature, and it only led to contradictions
in his exposition. Light is a medium for vision, 
never an object of sight. Observers can observe light but not by means of
reception of light and consequently not by means of sight. 

There are different classes of observers in special relativity and this is
the first reason for the non absolutism of time.
For instance, there are classical, semi-classical, 
and non-classical observers, all coexisting in the
relativistic experience. Relativistic phenomena
are integrations of different kinds of phenomena.
This is the same as to say that relativistic observations
are integrations of different kinds of observation. To be more precise,
a unique atomic object emitting light can be considered
as a clock or as an indirect means
for observation of material processes 
or movement of bodies, but it can also be taken as an objective material
event, independently of the light which it produces.
This means that different messages arise from the same
set of numbers of cycles of light. The point to be kept in mind is that
if light is a means for observation, it cannot be objectively observed
considering a single observer, these two roles are antagonist.
It is based on this kind of antagonism that different
classes of observer have to be defined.

So, it is not because observers are 
inside or outside the train that they find classical, 
non-classical or relativistic results. 
All these misunderstandings about observers, sensors, perceivers, thinkers and
experimenters, only show the necessity of a fundamental revision of the 
concept of observer. Observers are beings in transformation, inside a
developing physics.  It is not only in relativity that we find 
observational changes. 
Physics developed in the direction of more sophisticated theories,
but the physical world became somehow detached from the experience 
of the physical world, expelling the observer. It is a common
saying in the physical environment that 
physical phenomena can be derived from abstractions, such as
Minkowski space, wave functions etc. These abstractions
are not considered to be physical phenomena and, consequently,
they are not physical entities. Not worrying with the fact that, not being physical, 
they still must have a nature, many scholars do not realize that in this way they
can be deriving physical results from the metaphysical world. So, if we do not find
the observers of the physical worlds, it becomes very difficult to explain 
physical phenomena. Without 
physical observers the world has no physical meaning.

The organization of this paper is as follows.  
Section II is about the several concepts of observer and Einstein
experiment.
Section II-A discusses different kinds of
observation comparing perception with
physical observation. 
Section II-B recalls the setup of the train/embankment 
experiment and its main points concerning 
the present work. 

Section III is about the observations of Einstein experiment. 
Subsection III-A is about the basic observational concepts.
Subsection III-B describes the conditions of the observer who,
in the opinion of Einstein, should be inside the 
train. 
Subsection III-C describes the observations of a person 
who was considered by Einstein as being on the embankment. 

Section IV is about observation depending on reception of 
waves. 
Subsection IV-A discusses common points in observations 
depending on reception of material waves and
on reception of light waves. 
Subsection IV-B shows that different kinds of
observation correspond to 
different concepts of velocity for light.  
Subsection IV-C explains how different measures 
of time can arise for different observers.
 
Section V is about relativistic observation.
Subsection V-A comments on the origins of the equivalence of Galileo.
Subsection V-B introduces a generalization for the Galileo equivalence. 
Subsection V-C introduces a principle of conservation of information.
Subsection V-D finally presents the derivation of the constancy 
of the velocity of the light. 

In the conclusion, we add a with few comments 
about the relativistic observer. 

And in the Appendix we present 
a model of observer, based on the ideas of 
Jung, Weyl and Carnap, which underlies our reasoning.

\section{About observers, receivers of light and perceivers
in the experiment of Einstein.}
 
\subsection{ The various concepts of observer}

Observation is a process of description. 
There are many kinds of observation and consequently of phenomenon. 
Physical observers correspond to physical phenomena, and are 
expressions of specific
physical languages with their own concepts and expressions. Psychological 
observers also exists and they are responsible for the 
psychological phenomena, described by means of their specific language too.
Perception, or observation of the senses, is  not physical 
phenomenon but psychological event, although it depends on localization of observers 
as physical bodies. Because a
sensorial experience originates from the interaction of
a being with the environment, perception depends on the
spatial localization of the observer's body. Vision and hearing, among
many others, are modalities of perception. And, because two physical
persons are not identical, and neither can be at the same place at the same time, 
perceptions are individual and subjective experiences. 

Vision, which is one between the several modalities of perception,
is the result of a combination of many processes of different nature.
It is a psychological phenomenon but it also has a physical base. 
Real vision only happens
with the physical stimulation of the retina and consecutive 
processes in the body of the perceiver. The retina is a structure of cells with
a layer of photo-receptor cells. From the physical point of view, we may assume that
space is filled with a non-homogeneous spatial and temporal distribution 
of matter, which produces, absorbs and reflects radiation. In this picture 
we also can consider the existence an open and finite surface 
layer (which can be a retina) of a certain
body inside the space, also considering radiation constantly 
reaching the surface. Light reaching this physical retina 
produces a distribution of energy inside the material of the retina, 
corresponding to a radiation pattern reaching the surface. 
This superficial pattern is due to the luminosity
which results from the distribution of sources outside. Then,
there is a correspondence between the
energetic distribution inside and the distribution
of matter outside. The resulting processes inside
the material retina are propagated to the interior of the body
and these processes all depend on the kind of matter 
and material structure of the first layer and interior
of the body.

If the physical outside can now be considered only one, the retina
has two functions. In sensorial terms, the retina
becomes a superficial distribution of brightness.
While luminosity is purely a physical concept,
brightness is a sensorial experience.
Psychophysical measurements,
relating reports of people to physically controlled stimuli,
account for relations between brightness and luminosity.
The perceptual world of vision starts with distribution of
brightness and ends with distribution of bodies in
space. But these are not physical bodies or a physical space.
The contrast in brightness is one between the 
sensorial and non-sensorial elements
forming the perception of individual bodies
and emptiness that is interpreted as bodies in space.

In spite of the fact that perception of space is not only a visual 
acquisition but has other origins, visual perception  
always includes space and it is 
not yet completely understood. With spatial perception, a geometrical 
configuration emerges and
the experience of a distant object occurs, 
which actually does not coincide with the physical picture of
the facts. It is well known from psychophysical experiments that perception and physical observation
do not share a common geometry. This fact was firstly realized by
Rene Descartes, who was the founder of the science of psychology. 
Before him, psychological phenomena were usually assumed to be manifestations of a religious soul.
He inaugurated a revolutionary approach of the psychological phenomenon, by stating 
that soul manifestations and physiological processes were correlated.
Correlations of this type form the science of psychophysics and, until nowadays,
this empirical science is the only available tool to study the neuro-psychological 
correlations. The question of the space can also be studied by 
means of the psychophysical approach and many research had been done on the issue
of the perceptual geometry~\cite{Koenderink2000}.

Although the notion of space, as emptiness,
seems to be a single and innate idea, perceptual space and physical 
space are not the same concept. 
Then, we can assume that these two spaces, the perceptual and the
physical, are separated from the beginning. 
This means that when we say that we see the outside,
we are only speaking about the perceptual outside which, in the
end, is a representation of internal processes of our body. The
physical outside can be known but it cannot be seen.

Vision is only one among the several kinds of observation 
based on reception of light. 
The retina is just a material register, like all
physical registers and, due to this fact, it shares
with all registers the property of not being able to
account for processes which are not registered. Whatever can 
be registered, only becomes registered by means of the matter of
the register. All this gives us the certainty that we are not 
able to see what
is physically outside, as light waves must be, in order to
make our vision. The concept of light wave is a pure physical concept
and, if we conceived it, it cannot be because someone saw it. 

Contrary to perceptions, physical observations are
objective experiences in the highest degree. 
Experiences of a single person do not play any 
role in physics, what one single person observes is what everybody in all 
places, even moving with respect to each other, can observe. 
Therefore it is clear that a physical observer cannot be 
a person perceiving.
To really understand physical observers, it is necessary to know
that not everything we know happens to us because of the interaction
of our body with the external world. It is not only through sensations that
we know the physical reality, on the contrary \footnote {We are able to describe 
objective experiences 
not produced by any kind of stimulation triggered by the effect of  
the physical
outside on our sensorial organs. With this we are not saying that
thoughts can exist without brains. What we say is that there are thoughts,
occurring concomitantly with the stimulation of the neural system,  
which are not originating together with sensation and perception. This is not at
all an original assumption, it is well known in neuro psychology. 
But the main interest of the sciences influenced by
the logical positivism, dominating the last century, was to deny the existence
of this kind of thought. Special relativity
was the main representative of these sciences.}. It is a very known fact 
that we humans, not only are able
to experience much beyond our individual circumstances, but 
we are also able to know aspects of the physical world without the help 
of experimental proceedings. The history of 
Knowledge is full of these cases, the ideas of  
rectilinear uniform movement and the existence of atoms
are a few examples.

In the context of Einstein's experiment, the only common point 
between perceivers and the physical observers, of classical physics, 
is that both are
receivers of light. By definition, a receiver of light
is a being that only observes or knows about processes or 
events producing light, by means of light emitted from them. 
Receivers cannot observe the light traveling in space, 
because for this, they would have to receive (or register) light
instantaneously from the complete light wave spread and traveling in 
space. This cannot be, because this would imply that two different kinds of light, 
one traveling with infinite velocity and another with its usual 
velocity should be present together. As a consequence of the fact that
they do not observe light traveling in space, they do not know 
that they receive light to observe far events, they only observe these events. 

The main difference between a
classical observer and a visual observer is in which way distances
are defined. In both cases, registration of light is local. In 
case of visual perception, distances are not
necessarily quantified and when it happens, quantification
does not mean an objective proceeding. This means that distances
are subjective and locally defined. But in the physical
case, distances are defined by physical means, independently of
reception. And an observation is not a registration but a continuous
association between registrations and objective distances, making
the physical observation a non-local phenomenon. The objectivity
of the distances comes from the physical knowledge of the
rectilinear uniform movement, which is an {\it a priori} element
of classical observation. Due to this fact, even for receivers
and classical observers within relativity, 
observation is nonlocal since distance is nonlocal by principle.

But this is not all about physical observers. When light reaches a register, 
it can be interpreted as a reception of light if one accounts for the fact 
that a wave of light 
was emitted from a source, reaching the register. 
But it can also be interpreted as a material process 
of a distant object, when one does not account for the travel 
of the light wave. 
The first case clearly implies the existence of another kind of observer, the one 
who is able to describe the movement of the light wave, without reception.
 
With the observation emerging from sensation, we have perception,
and the perceptual concept of brightness. In the physical domain
a new kinematic quantity, characterizing reception, must be defined. 
This cannot involve the concept of luminosity,
which in the end, is a dynamic concept, measuring
the effect of the radiation
on a surface register.
But considering that the retina can register
oscillations of brightness, we may consider that in
the specific case of measuring numbers, perception and 
physical observation are coincident. In analogy with
the perceptual case, we may assume the concept of physical
brightness, represented by the number of waves reaching
a detector or retina, as an observational feature of the
physical object.  
In this sense we may say that our body can also act as 
a physical register, having other means to 
observe, in the 
physical sense, the source of light. Otherwise, we cannot
be sure that there is indeed a source somewhere else and 
the sensation of brilliance cannot be taken as objective.

The term ``physical brightness" is only an analogy to the perceptual
experience of bright, since the spatial 
localization of persons does not influence the physical observation.
From a device receiving light, the number of received 
waves would be the signal of the existence of the 
oscillatory object at distance. Brightness here is just a name for
what becomes the signal, or representation, of the object. 
The possibility to connect these two distant realities, does not
imply in the physical localization of a person, but
implies the existence of a person. Actually, this
connection is the mark of the human being: the fact is that
we make these connections. Here we do not
explain them but only take them for granted. Moreover,
we also can avoid them, and it is exactly what we do when
we interpret the 
same data, from the same device, 
as a pure local event. And it is not a question of our physical
presence whether we tell one or another story about the same data.

So, Einstein could also have considered a large crowd
spread on the embankment as well as a limited train filled with
many beings. The measurements in the embankment and the ones 
in the train could be known
by each and every one, independently of being in the embankment or in the train. 
Neither 
was the movement the reason for the observational difference. 
What actually produced the
difference was the way in which data made by light were interpreted in
the description of the facts.
The physical world, that is, the 
set of all physical experiences, is not
the perceptual world of our daily experiences, the
one we can see, smell and touch, as people usually
suppose, but the one we know by means of 
physical observations.
Physical observers \footnote{ The expression `physical observer' does not mean
`observer with physical body' but `observer of the physical world'.}
are not beings to be found inside the physical space, 
measuring or doing something else.
Actually, this observer of the physical literature is 
just a name given for a special kind of
thinking representing a specific collectivity.

\subsection{Einstein's train/embankment experiment} 

The train/embankment experiment was introduced by Einstein to explain special relativity. 
In essence, the situation discussed is the following: 
there is an embankment and at a certain initial moment $t=0$ the origin $\Ot$ of a train passes 
through the origin $\Oe$ of the embankment,
with a constant velocity $\ve$. At $\Oe$, there is a source of light. 
\footnote{\label{Albertslamp} 
Though we shall follow Einstein in considering light, our discussion would be easier 
to follow when speaking about the equivalent case of radio waves, because then the wavelengths 
exceed the size of human bodies, etc.}

An observer\footnote{We repeat Einstein with the idea of persons as observers
otherwise we cannot discuss his experiment. But the reader can always 
think of many persons describing the same observations}, named $\obst$, is 
located in the train and employs its three-dimensional 
walls, with origin at $\Ot$, as his inertial reference frame. Another person, 
named $\obse$, is at the embankment, with origin $\Oe$,
using it as his inertial reference frame. It is said that 
because the velocity of the train approaches 
the velocity of light, relativistic effects are noticeable, between the two observers, 
such as discrepancies in measurements 
of space and time. 

But, in his experiment, $\obst$ is a genuine classical observer, measuring 
his `proper time'. Consequently, he cannot know that his time 
is different from the embankment time and neither can he use the constancy 
of the velocity of light to know where the moving object is, because this
is a condition of special relativity only, not of classical physics. It is important to notice that, 
in Einstein's exposition, the embankment has no limit in size, the experiment has no limit in time and 
the embankment clock serves the whole embankment. This means that $\obse$, describing all the sequence 
of positions of the train, uses only one clock. Consequently, for this observer there is no sense 
in the concept of local time and he has no reason to suspect that inside the train there
could be another time, not even if he is able to read off a clock moving with it.

Based on conventional ideas about what observation is\footnote{The assumption that 
to observe is to measure.}, we assume that observers only measure distances
and intervals of time. Trying to understand this, we assume that we only 
can be in the conditions of one or another observer. 
And, because both do not understand each other, we finish 
not understanding the relativistic reasoning. One of the problems 
in the exposition of Einstein is that it 
does not explain the true origin of the non-absolutism of time, that is the
observational difference. Because of this, he does not tell us that the relativistic
conclusions are only due to the existence of a specific observational entity, the
relativistic observer, who defines his own relativistic concepts and whose
conclusions are coded in the equations of special relativity.

The non-absolutism of time is the first serious discrepancy introduced by special relativity.
Discrepancy means difference or disagreement, also implying comparison or a certain level of equivalence.
In our opinion, there are discrepancies in measures of lengths and time, first of all, because the two 
observers are different and remain different. They do not know the discrepancies and are separated 
from the beginning. But their measures can be compared by someone else, in other conditions. Time
is non absolute only for this third observer. 

In our experiment, $\obst$ measures time from a near clock but depends on
reception of light for observing objects at distance. But he cannot observe the
light waves he receives. We will show that it corresponds to the situation of a classical observer. 

The other observer, supposed to be on the embankment, 
measures time from a clock at the embankment, 
observes the movement of 
the train, the wave of light in the usual three dimensional space, but he does not need reception of light 
for observation of light. This one is named the ``semi-classical" observer. 

When the two origins are at rest 
together, they agree with all their measures of space and time.  However, when moving 
away from each other, these measures do not coincide. 
In all what follows, we make the hypothesis that light as a wave phenomenon, consists of a series 
of pulses in space. To simplify matters we consider monochromatic waves only.

\section{The two types of observer}

\subsection{Physical concepts and observational elements}

We find in the physical literature, concerning relativity, distinctions in concepts of time such 
as proper time and relativistic
time. The proper time is usually defined as a local measurement of time, not depending on optical 
means. It is used in this sense that light does not convey information from the clock to the observer. 
But the so called relativistic measurements, are also measured by a local (in the same sense) clock, 
usually serving the whole space of the experiment. This is the case for relativistic events inside
laboratories. Only one clock on the walls can be used to measure the so 
called relativistic time. Therefore this distinction, based on
distance from the clock to the observer \footnote{Frequently in the literature, the authors think 
of observers as
beings with bodies, who can measure and disappear when convenient.}, does not 
give a meaning for proper time. 

As we explained before, an observer is not
a localized physical person perceiving events and making the measurements, 
but the one who describes
events, by associating data 
\footnote{In a physical sense, the observer 
(when considering the body of a human being) can be 
replaced by a photo camera, even read off by a computer program, 
as is standardly done in high energy physics. But somewhere down the line, someone has to 
organize and interpret the data, making up the event by integrating the concepts. This final result
is what expresses events such as uniform motion, 
movement of light, etc.\label{fncamera}}. 
In general, a phenomenon (an event in development, or just an event) is a continuous succession 
of happenings in time. A basic feature of physical events is that time is a
quantitative reference for them. 

Time is quantified by means of units or cycles.
Cycle is the common feature of all devices named clock. The cycle is a unit of time, 
and, for this reason, it is not a spacial entity.
However, clocks can have spatial dimensions.
A clock is a reference for time, a physical event with the property of being cyclic. There 
can be several references of time, based on the
same unit of time. The observation of a physical event is based on an
association between at least two events, one of them being the clock. 

A wave is another entity which has a spatial dimension, it cycles and moves 
in specific directions.
The origin of a wave is a cyclic event, independent from the 
wave itself. As a pure kinematic entity, it combines a spatial dimension 
to cycles, making moving pulses, transferring cycles from place to place.
But the description of a wave is based on the usual references, including a clock,
independent of the wave. The pattern of cycles carried by the wave can vary
with respect to moving references and can be compared in different places with respect to the same
fixed pattern of cycles of the original clock.

Differences in measures of cycles of a wave due to movement, 
are called the Doppler effect.
The difference between the so-called classical Doppler effect 
and the relativistic Doppler effect lies in the uniqueness
of the original pattern of cycles, generating the waves.
In the non-relativistic case, the observation of these original
cycles does not depend on reception of waves while in the relativistic 
situation, due to the differentiation in the
classes of observers, this original pattern is not 
observed independently of the reception of waves. Then, an extra
correction in the observation of the cycles must be made.

Now we study the different ways of observing the movement of sources and
waves and compare them.
In our discussion,  
the origin of the embankment is $\Oe$, where there is also an atomic clock 
\footnote{An atomic clock is a high-stability oscillator based on atomic transitions. 
The second is standardly defined 
as the duration of 9,192,631,770 cycles of microwave light absorbed or emitted by 
the hyperfine transition of 
Cesium-133 atoms in their ground state, undisturbed by external fields.
In this definition we find the word duration. But `duration' is not something which exists
{\it per se}, without the physical object which defines it. This is exactly the same
 as in the case of lengths, in which we must have the physical object to be the standard
measure. The length is the object in the same sense that the duration is the object too, in
this case the specific atom. So, the word duration makes the definition 
circular and should be omitted.},
from which electromagnetic pulses originate. 
This clock is a cyclic event of the same nature as a lamp 

There is also an origin $\Ot$ inside the train, in which another clock is localized. We also consider that at a 
certain moment $t=0$ the point $\Ot$ of the train coincides with $\Oe$ of the embankment. 

In other examples with two observers, Einstein assumes that each observer must have a lamp. We consider lamps 
and clocks as identical objects, concerning their descriptions as cyclic phenomena. Atomic clocks 
also emit light. A lamp is just a name for
speaking about an atomic process emitting light and for this reason it can be used as a clock or vice versa. 
In our discussion, the cyclicity of these objects is their most relevant observational feature. 

Our interest is to study each observer, how he counts 
cycles or pulses of light, how he interprets 
them and what light is for him, pointing to the differences 
in the ways of observing. We are going to put forward
that these differences are, in fact, the origin of the non-absolutism of time.

\subsection{The observer in the train}

We start by assuming that the observer in the train, $\obst$,  
is a classical observer\footnote{We could alternatively 
assume that observer 2 is a classical observer, 
with the same results.}. 
This just means that he only describes the movement of 
the embankment, or of an origin in the embankment, 
with respect to his inertial frame of spatial references at rest with respect to the train. And this 
is done according to the time measured by his clock inside the train \footnote{What we mean with 
`classical observer' cannot be fully explained at this moment. 
Our method is to state the classical conditions from a non-classical but not yet relativistic
context of observation.} 
 
At this point we just assume that, if this observer were alone, classical physics would 
be the context of his observations.

About $\obst$, we state the following conditions:

1-1) He is able to count the number of cycles;

1-2) He uses a clock to measure time;

1-3) He is a classical observer. Consequently, his only expression has the usual 
shape of a rectilinear uniform trajectory: 
\BEQ \label{1} \st=\vt \tt \EEQ 
where $\st$ and $\vt$ are the position and velocity of $\Oe$, respectively, 
with respect to the train, 
and $\tt$ is the time in the train measured e.g. using his Ce-clock at $\Ot$.

\subsection{The observer on the embankment}

The second observer, $\obse$, describes the cyclic event at the origin of the embankment, the transmission 
of the waves from there, and their detection in the train.  There is a clock at the origin of 
the embankment. $\Obse$ thinks of
the embankment as being at rest and also describes the movement of the train. For this observer, light is
an objective event, happening concomitantly with the other ones, but it is not a means to observe.

The crucial difference between this observer and the previous one is that he can describe light 
objectively. This means that he knows where the light wave is in space, without the need of detecting it, 
to observe it. Experiments may have been done previously, in order to give him the certainty of the velocity
of the light with respect to $\Oe$. He can also arrange things in order to know, in advance, all 
the times and positions from the beginning to the end of the experiment. But, in this experiment,
he does not receive (detect) light, from any point and moment of the light trajectory, 
in order to know where it is. This would be in complete contradiction with the fact that the 
velocity of light is finite. Light
is not and cannot be a means for observing itself, in any case. In 
this sense, an observer of light is not classical 
because, in classical experiences, objects at remote distance can be known without
considering reception of light, what means without knowing that
reception may play a role on measures of time and consequently of distances. A 
classical observer who receives light is not aware of this condition, and cannot know light as it 
is. For reasons which will be more clear later, we assume that $\obse$ is a semi-classical observer. 
\footnote{ A classical observer cannot observe light by 
definition. An object with the properties of an electromagnetic wave does 
not obey the Galilean 
rules and consequently is not classical. It is important to realize that 
expressions such as `classical 
aspects of light' are just ways of speaking, without correspondence with 
any physical language. But, since we describe
the movement of a wave light, we have to assume that we can also observe 
non-classically. This is what we assume here. }

2-1) The equation 
\footnote{\label{vsign} To avoid complications in notation,
we assume that $\obst$ and $\obse$ count their lengths and velocities in opposite directions, such that
$\st$, $\vt$, $\se$ and $\ve$ are all positive.}
\BEQ \label{2}  \se=\ve\te \EEQ 
is also his form of the movement, where $\te$ is measured according to next definition, 2-2).

2-2) At $\Oe$ there is also a source of light or a lamp. The production of light 
is described by a frequency that is defined by an association of a cyclic event with 
a clock, which is another cyclic event. Frequency is the number of cycles per time. 
In this way defined, this is not a fixed number but counted as time goes by, 
therefore it is numerical function of time. 
Without the two events being compared, neither the number 
of cycles nor the frequency can be measured. 

For $\obse$, the original cycles of the atomic event are described by means of a 
time reference that is also based on production of light from cycles. It is possible to
define a generic standard unit of time $u_l$, by means of the emission of a light 
pulse according to chosen material conditions and specifications \footnote{The usual 
second is defined as described in footnote 10. },

The original cycles in $\Oe$ are atomic transitions with frequency: 
\BEQ \label{3} \fcyoee=\frac{\Nc(\te)}{\te}, \EEQ
where `cy' stands for `cycles', $oe$ for origin of the embankment and $e$ for  
$\obse$. 

In this equation, $\Nc(\te)$
is a pure numerical function, growing linearly with time, 
describing the number of cycles during an 
interval of time $\te$, as counted by the observer. 
Because time for us is 
an ordered succession without end, 
this expression does not mean just a fixed number but a process of successive 
numeration, considering a specific interval of time. For $\obse$, the original event 
in $\Oe$ is observed by counting cycles and measuring time. 
 
2-3) In principle, the light wave is a spatially extended 
phenomenon, distinct from the
material object producing the pulses. 
The frequency of the wave is given by 
\BEQ \label{4} \fwoee=\frac{\Nw(\te)}{\te}, \EEQ
with subscript $w$ denoting wave.

It just holds that the number of cycles made by the lamp equals the number 
of cycles of generated light, 
$\Nw(\te)=\Nc(\te)$, so, in this case, we also have the
same frequency:
\BEQ \label{5}\fwoee=\fcyoee. \EEQ 
As it is well known, a wave is not completely described by a frequency, 
but also by the wavelength and the amplitude. While a cyclic phenomenon can be localized in a point, 
a wave is an extensive object. The distance reached by the wave front in the embankment frame equals 
\BEQ \label{6}\roee=\ce\te\label{r=}\EEQ 
where $\ce=c$ is the velocity of the light wave front with respect to the embankment. This velocity 
can be given by 
\BEQ \label{7} \ce=\frac{\Nc(\te)\lambda_e}{\te}, \EEQ 
where $\lambda_e$ is the 
wavelength. Measuring $\ce$ in the embankment frame, the wavelength can be found from: 
\BEQ \label{8} \ce= \fwoee \lambda_e. \EEQ

2-4) We assumed that $\Obse$ is not a classical observer but he is also far
from supporting a relativistic result: he can observe the relative
movement between the wave front and the train. 
He knows that the 
velocity of the train is approaching the velocity of 
transmission of the pulses and that, 
for this reason, there is a decrease in the frequency of the wave measured at $\se(\te)$, 
given by  
\BEQ \label{9} \fwote=\frac{\Mw(\te)}{\te} \EEQ
where $\Mw(\te)$ is the number of cycles at the origin of the train.  
$\Mw(\te)$ and $\Nw(\te)$ are different numerical functions 
of $\te$, $\Mw(\te)<\Nw(\te)$  
for every $\te$.

$\Obse$ can describe the velocity of the wave front relatively
to the train by means of the speed
\BEQ  \label{10} \cote=\ce-\ve, \EEQ
where $\ce$ is given by equation (8), 
and $\ve$ velocity of the train. 
It is important to notice that, for this observer, the relation  (10) 
is an initial condition which does not change with respect to the increasing 
distances to the train.

And, according to this semi-classical
observer, this velocity should be given by 
\BEQ \label{11} \cote=\fwote\lambda_e, \EEQ
Substituting equations (9) and (11) in equation (10), we find
\BEQ \label{12} \fwote=(1-\frac{\ve}{\ce})\fwoee.\EEQ
This frequency, on the left side of the equation, is a concept
made by associating two sets of measures, the sequence of times and
the sequence of numbers $\Mw(\te)$. Both sequences happen
independently for $\obse$. He measures the numbers $\Mw(\te)$ just 
by counting cycles arriving at $\Ot$. 
We will show that $\fwote$  is not measured in the train.
But we will also show that the sequence of numbers  $\Mw(\te)$ is
measured by $\obst$. 
The relation between frequencies can be taken from 
\BEQ \label{13} \Mw(\te)=(1-\frac{\ve}{\ce})\Nw(\te).\EEQ

Equation (12) shows that, in spite of observing light,
$\Obse$ is limited to the clock at the embankment and he can describe 
the situation as a normal classical Doppler shift. 

This result is not completely
classical because a classical observer would not observe light waves.
But it is also not relativistic, since it was deduced from a limit in which
there is a direct (classical) association between the sequence of cycles 
$\Mw(\te)$ and the sequence of cycles of the clock at $\Oe$. 

Equation (13) does not depend on equation (12) but derives directly from 
equation (10). This last relation is 
integral part of the relativistic
description and the origin of factors like $(1-\frac{\ve}{\ce})$ or $(1+\frac{\ve}{\ce})$,
resulting from relative velocities.

\section{Non-classical analysis}

\subsection{Considerations about the two observers}

To say that a wave is observed in space and its behavior described, is the same as to say that
all regions of the space filled with the undulatory entity are known simultaneously and that 
this instantaneous information is continuously received, or continuously kept in some way, 
during the period of observation. 
This is not only for waves 
in space but it is true for every extensive body and
objects of any kind, such as a distribution of numbers, matter and so on. The word `simultaneity' is used 
for this property of global observations, associating two or more regions of events, or even 
non-spatial events, with one time. It implies a diversity 
in the associated objects, without implying any condition on the physical nature of these associated  objects. 

Going back to our situation, there are oscillations being counted in the train. According to $\obse$,  the
frequency of oscillation is given by equation (9). We can reconsider this, picturing the positions of each object 
in the embankment time and focusing our attention on $\obst$ 's measures and thoughts. It is true that he knows 
very well the sequence of numbers $\Mw(\te)$, 
registered on his apparatus or detector. But he cannot 
agree on the same relation (12) with $\obse$. 

Equation (12) is known to be a relation between frequencies of the same object (the wave) observed under different 
conditions with respect to movement. $\Obse$  knows the whole extension of the wave, in the sense explained above 
and consequently he observes
the relative movement between the train and the wave.
But $\obst$ lacks this knowledge of unity of the wave and even of the wave itself, which is so natural for
$\obse$ .  This is not due to the nature of the wave, in first place, but to his condition as observer 
depending on the reception of the waves to observe far objects. 

One single wave cannot convey its own pattern of distribution in space. 
This object we name `wave' exists between different regions of space and may grow in distance
but it does not transport itself between these regions. We mean that a wave conveys something else, 
not information about itself, except if its existence is already known independently
of detecting its oscillations. In order to know the pattern of distribution of
a wave it is necessary to have another wave, or other means of knowledge. And this holds
for all kinds of waves. 

Exactly like a sailor in a boat, $\obst$ measures the cycles of the wave at
$\se(\te)$. But the observation of these cycles in the boat (only up and down) does no 
inform him about the 
distribution of waves on the water,
far from the boat, 
already settled at each moment he measures the cycles. Unless he could know the wave
distribution on the whole surface of the
water, by other means much faster than the wave whose cycles he measures, 
he cannot know the
relative velocity between the boat and the wave and neither the frequency 
of the original source
$\Oe$.  In the end, he only knows the number of his local cycles. 
Equation (12) again does not apply to this case.
Implicitly in this equation is the simultaneous knowledge
of the different regions of the wave, and consequently their different frequencies, or the relative movement 
between the wave and the moving reference $\Ot$. 

But we know that waves on water are very slow in comparison to light. All the problem of observation 
of the water waves would be solved by reception of electromagnetic waves, giving very fast (practically 
instantaneous) information
about the positions of the source or the traveling pattern of waves on the water and consequently of
the velocity of the water wave. In the absence of another wave, the observer in the boat would have only 
the sequence of cycles of his apparatus to consider. Being used as a clock, the measure of received
cycles would represent another notion of time\footnote{This situation applies for organisms such as 
jelly-fish, if this animal is not capable to use his biological processes as a clock.}.

For the case of electromagnetic waves as the faster medium, exactly the same could happen. 
The electromagnetic wave is 
produced by a cyclic atomic event, also used as a clock, and it is the faster known medium which can 
convey the information of these cycles. The atomic cycles which appear in the measuring apparatus 
of a moving reference, correspond to the original cycles. But, in the absence of another process,
these same cycles must be used as a clock, representing a new unity of time. For someone 
counting the cycles from his local apparatus and also 
making from it a reference for measuring time, in the absence of another reference, there 
appears a difference in the measurement of time as well, with respect to the original measures. 
The problem is that observers in these situations do not have means of knowing their mutual differences
in outcomes of measurements. As soon as the spatial frames of reference separate from each other, they loose
information with respect to time and the only means of communication takes time to reach the other. 

These are situations which may happen but they do not fall under the 
description of equation (12). Observers depending on these moving references, would not
have complete information about physical facts, such as the processes which generate the waves. 
This does not depend on the kind of wave but only on the dependence of the observer on the wave.
In the case of light, it would be a lack of knowledge about the atomic processes producing light, 
independently of its propagation,
what actually would mean subjectivity. This is a condition which cannot be 
accepted. 

The difference in time arising from this situation 
can be expressed by a proportionality between the two times. But these different 
measures of time do not necessarily 
mean an independence of the references, in 
the sense of a genuine new dimension of time, to be named non-absolutism of time. 
We must reserve this term `non-absolutism' for a situation
in which we can be sure about the objectivity of both observations, what is not yet the case
of the situations above described. 
The fact that moving references exist, with speed comparable to the speed of light, 
requires a solution for the problem of objectivity. 

\subsection{The various concepts of velocity of light}

In the previous analysis there was no remark about what $\obst$ knows 
about the train itself,  
which remains the same entity as time goes by. Since the train
is a material object, we have to assume that
$\Ot$ consists of atoms. In principle, this system of atoms does not interact
with $\Oe$, neither is it influenced by the 
existence 
of the embankment in any way. Emitting light, these atoms 
in transition also form a cyclic object.
We can assume that, with respect to this inertial system, the light emitted
from $\Ot$ moves away from it with velocity $\ct$.  

Now, to avoid misunderstandings, we recall all possible 
velocities of the waves until now involved in the discussion.
In our new notation, the inferior indices represent the origin and the
reference of the velocity of wave, respectively. The superior index
refers to the observer. We have:

$\coeoee=\ce$, the velocity of the light emitted from $\Oe$,
relative to $\Oe$, according to $\obse$. 

$\coeote=\ce-\ve$, the velocity of the wave emitted from $\Oe$,
relative to the train, according to $\obse$; 

$\cotote=\ct$, the velocity of the wave emitted from $\Ot$, 
relative to $\Ot$, according to $\obse$.

$\cotoee=\ct+\ve$, the velocity of the wave emitted
from $\Ot$, relative to the embankment, according to $\obse$.

The fact that $\Ot$ is a cyclic event can be known
by $\obst$. And, for what matters,
we do not need to consider extra Cesium atoms
coupled to $\Ot$, because the relation between the cycles of 
both would be fixed from the beginning and it would not add extra
information about the cyclic state and the movement of $\Oe$, 
for $\obst$. We just consider
$\Ot$ as a material sample, identical in all senses to
$\Oe$, not 
influenced by the light wave from $\Oe$. This
implies that $\obst$ can share with $\obse$ the 
same number
$\Nc(\te)$, when observing $\Ot$.

According to $\obse$, 
light from $\Oe$ reaches the position $\se(\te)$ of the train with 
velocity $\coeote=\ce-\ve$ and
light produced in the train moves with velocity $\cotote=\ct+\ve$, with respect
to $\Oe$. 
That these light waves produced in the train may leave it with velocity 
$\ct=\ce$, with respect to the train,
is not the point of our discussion. 
We assumed an $\Ot$ physically identical
to $\Oe$, in all senses. The real point of our discussion is to understand the 
relativistic claim, which prohibits the existence of relative velocities of light
with respect to moving detectors or sources. 
Special relativity does not say anything about the reason
why the value of the velocity of the light {\it in vacuum}, 
with respect to its source at rest, is $c=299,792,458 meters per second$.
Consequently, we must assume that $\ce=\ct=c$,
since these are the concepts of velocity, defined with respect to
their corresponding sources at rest.
Instead, the theory says that light 
from $\Oe$ reaches 
the train with velocity $\ce$ and the light from the train, which in
principle should be independent from the one
made at $\Oe$, also leaves it
with velocity $\ce$, with respect to $\Oe$.
According to the axioms of special relativity,
the concepts of relative velocity of light {\it in vacuum}, 
here represented by $\coeote$ and $\cotoee$, do not exist and, consequently, 
$\obse$ does not exist or is mistaken.

Usually, people interpret the facts as if $\obst$ could observe light 
arriving from $\Oe$ with velocity $\ce$, but, according to our interpretation, 
it is not $\obst$ who arrives at such a conclusions, and neither is $\obse$ 
mistaken.  
Because $\obst$ is not an observer of light, his registration of 
the wave coming from $\Oe$ has not the meaning of the arrival of a wave
but it has another interpretation. And in the case of $\obse$, we must
consider the fact that without considering a registration
(or reception) of the wave at the train, there is no meaningful 
relation between the movement of the wave and the movement of $\Ot$ 
which could justify a change in its velocity. For $\obse$ alone,
the increasing difference between the positions of the wave front and $\Ot$ 
suffices for an observation of a relative velocity between them. 
There cannot be a transformation in the velocity of the wave inside 
the embankment, only due to the fact that it shares the space with 
a moving body, with which light does not even
interact. Then, in our opinion, 
$\obse$ is free to keep his way of thinking. Something else must happen, and 
we focus on the
meaning of this light for $\obst$. 

With these same waves arriving
at the position of the train, another kind of observation
can happen. Because a wave can be broken in 
independent pieces of information, considering its features separately,
they can be combined differently, showing something
completely diverse. We need to understand what is the meaning of 
the arriving light for $\obst$.

From one point of view, we know that light
reaches $\Ot$, but from another this is not necessarily the case. 
For $\Obst$, who knows nothing 
about the spacial 
existence of the wave, 
what is out there is a bright $\Oe$. This is a conclusion which we take only
by analyzing our own ways of observing, as classical beings, and
comparing them with the situation here present. When observing, we do not
say that light is in between. 
If we drown in the condition of the classical observer, 
we have to conclude, liking or not, that light waves disappear as such
between $\Oe$ and $\Ot$. \footnote{ Until now, the expression
``bright object" could be exchanged by ``sounding object, "
considering material 
waves instead of light. We discuss the difference
later.}

In place of this light which disappears, another
concept of light must emerge, and this is in fact what happens with
the appearance of a cycling feature in the object.  
Reception of waves has a meaning by itself 
introducing a physical correlate of brightness, 
the kinematic brightness, 
which in this case, does not depend on the characteristics of the wave,
but only of the number of waves, as explained in the introduction.
And, for the receiver, the number of received waves is a feature attached 
to the objects at distance, re placing the absent wave. Here, this concept
only belongs to receivers of light, not to observers of light.
Reception generates a new concept of body,
the cyclically shining body, defining the receiver of light.
What we mean with reception of light is not a process independent from the 
observation, or causing observation. Reception is a process which happens together 
with the corresponding observation, furnishing part of the features of the 
observation. This cyclicity of the body, coming with reception,
is not a `thing' but a process, a body event,
having the duration of the observed event. 
So, reception is a physical process and a physical concept, 
involving physical elements, as
every physical event, but it is not
described by the receiver, who only gets a new concept 
of body from it.

The `brightness' has no physical 
relation or interaction with the cyclic $\Ot$, which is a reference for observing the 
object $\Oe$. It is by means of counting cycles of $\Ot$ and
combining it with the reception of waves from $\Oe$, that $\obst$
observes the movement of the body, which now becomes more than
the simple kinematic object from the pure classical physics. 
As we know from our own experience, if we think as classical 
observers, the final observation
could be just the continuous trajectory of a cyclically shining object, 
without any physical medium other than
the empty space between the points of the trajectory and the origin of the
system of reference. A pure classical body could also be described as 
moving object only, without necessity of any brightness. But here 
this feature is the new insight introduced by an
observer receiver of light, before the classical limit. His situation
is not yet classical, which is a particular case of receiver of
light resulting from additional conditions on the process of reception. 

As we can notice, light can actually have different meanings for 
different types of observer. While one describes the wave light, 
the other, by
means of the same light, observes
a bright body, which is a completely 
different concept. 
For $\obst$, light, as a wave, is beyond observation. Consequently, 
for this observer, light does not cross $\Ot$ because it never
enters his observational space, staying there with the body as 
an integral feature of it. In this way, the disappearance of one entity
is partially compensated by the emergence of another one. It is 
not completely compensated,
because brightness is local and the wave is not. Then, 
something is still lacking and we 
have to find the light which arrives and leaves $\Ot$
with the velocity $c$. Until now, what we know for sure 
is that if this light actually exists, obeying the 
relativistic claims,
it can only be
observation of someone else.

\subsection{The two measures of time}

Let us recall some aspects of $\obse$'s observations. 
Equation (13) relates two varying quantities, which are numbers
of occurrences or events,
by means of a constant. The number of 
occurrences and relations between such numbers can be 
physical observations, 
in the same sense that relations between time, lengths, velocities
and other quantities can be considered. Defining 
$\betae=\ve/\ce$, we rewrite equation (13) as:
\BEQ \label{14} \Mw(\te)=(1-\betae)\Nw(\te).\EEQ

We read in this equation that the relation
between the two varying quantities, 
is valid for each and every time. 
There is no specifications about the kind of the event whose number is 
growing in time, neither an indication
of a wave spread in space, there is only number of cycles
\footnote{For $\obse$, the existence of the wave is
stated independently by means of equations (8) and (11),
in subsection III-C.}. The name cycle here is used in the sense
that the occurrences are identical to each other. This
is implicit in the numerical description.
Then, we can imagine figuratively the whole phenomenon,
as a conjunction of two `closed' distributions of spots, moving between 
themselves and keeping increasing in number. 
In the observational context of $\obse$, the physical nature
of these spots is only specified by their number. This is a clear 
case in which counting is a physical description and number 
a physical concept. And we must
understand this fact directly from reading the statement of equation (14),
not from external arguments. 

Physical observations are expressions of the physical reality only, not
explanations about causes of phenomena. Kinematics suffices 
for the description and this is the reason why dynamical
quantities, related to interactions, and explanatory arguments are 
absent from observations. The kinematics here involves not only 
numbers but temporal conjunction of numbers. 
The existence of the number of occurrences without an indication of the physical 
nature of these occurrences, is very common in physical descriptions.
\footnote{The distribution function of events, extensively used
in physics of many-body system, is another example of a numerical concept
in physics.} 

Now, forgetting $\obse$, we think as $\obst$ observing $\Oe$, with respect to
the train, which has its own independent source of light $\Ot$ at the origin of his 
reference system. $\Oe$ is observed to
move away from it, as it happened for the previous observer. 
The difference now
is that $\obst$ observers the cyclic emissions from $\Oe$, but at 
the place of $\Oe$ by means
of reception at $\Ot$, while $\obse$ did not depend on local reception.
This is a condition of receivers of light and it implies that the 
number of cycles of $\Oe$, counted by 
$\obst$ cannot be the number of cycles of $\Oe$ corresponding
to time $\te$. 

Local reception has a proper meaning, which does not imply the
existence of someone
in the train or with particles, as we usually read on the texts. 
The meaning appears with observational statements, when we make
associations of far distances with local registers, when light 
takes time to reach the register. While building these associations
we make the meaning, without being specifically in the place of the register.
Since light actually travels with finite velocity from $\Oe$ to $\Ot$,
this takes time and what can be counted as cycles of $\Oe$ at $\Ot$, only can be
a previous amount of the cycles, considering the time $\te$.
This is the meaning of the quantity $\Mwt(\tt)$, instead of $\Mw(\te)$, in
the observational context of $\obst$. Now, the cyclic state of the object, 
which is a material process, is
measured by means of light and it fixes, objectively, the time of the receiver. 
The number of waves from $\Oe$ counted at $\Ot$ 
is the same for both observers, as we 
already said in the section IV-B, but this unique number
has different meanings, corresponding to the different observers.

We can write a similar relation,
for $\obst$, 

\BEQ \label{15} \Mwt(\tt)=(1-\betat)\Nctt(\tt),\EEQ

where

\BEQ \label{16} \Nctt(\tt)=\Nc(\te).\EEQ

With the introduction of this correction, we
can also assume that there is a certain factor $\gamma$ 
such that \BEQ \label{17} \te=\gamma \tt.\EEQ

For both observations, considering both observers,
the physical existence of those numbers of cycles is not to be explained
as a result of any kind of interaction. 
The pictorial spots can also be thought
as effects of the passage of an undulatory 
medium, without considering interaction between a register and the 
medium. We assume that this is not a physical effect but just an 
observational effect, in the sense that no interaction is involved which could
make of $\gamma$ a function of time. 
This is an argument to consider $\gamma$
as a constant.

This factor $\gamma$ is the mark of the observational
difference. The introduction of this temporal difference,
brings again the question of the objectivity. But, partially, it is solved with the
assumption that at $\Ot$ there is an independent source of light. The 
other aspect of the problem involves the observation
of the movement and consequently of the velocity, for
$\obst$. This problem has to 
be solved by shaping the final trajectory of $\Oe$
for $\obst$ and comparing it with the trajectory of
$\Ot$ for $\obse$. In other words, the 
solution of this problem depends on conditions in which
the receiver becomes a classical observer. 

Until now we find that the origin of the time difference
is not in different clocks but it is in
the different ways of observing. 
In principle, this would still hold if we exchange
the shining object by a sounding object.

\section{Special relativity }

\subsection{Relativistic arguments }

It is interesting to notice 
that the classical and semi-classical regimes of observation, of
section III are never subjective or wrong. 
The classical observation, equation (1), $\st=\vt\tt$, is objective and correct in all senses. 
It is so that, for $\tt=\te$ in quantity, it agrees completely with equation (2),
$\se=\ve\te$, from $\Obse$ language. And nobody can affirm that
$\Obse$ is wrong, by defining the relative velocity of equation (10), $\cote=\ce-\ve$. 
For every $\te$
there is actually a difference between the position of the wave front and the
position of the train such that equation (10) can be assumed as valid.

The problem is that equation (10) does not necessarily imply equation (12),
which describes a rate of emission of a wave light, considering a certain position.
Equation (10) and equation (11) are independent equations. Rigorously, the 
relative velocity of equation (10) 
and the velocity of equation (11), $\cote=\fwote\lambda_e$, are not the same physical concept. 
Inside the limited context of $\Obse$, this fact is not explicit. 

Equation (11) does not hold for the receiver because his measure of time is 
another and consequently the frequency is not the same. Therefore there must be
a general equation, instead of equation (12),
that integrates both observers. About a comparison between the two observations, it
is also interesting to notice that the observations of $\obst$ are not under the 
observational domain of $\obse$. A comparison implies the knowledge of
invariant quantities by means of which the two observations can be related.
Until now, equations (14) and (15) are completely independent. Since we do not 
know the quantity $\Mwt(\tt)$, we also do not know $\tt$.

\subsection {Postulate I: Generalization of Galilean equivalence.}

Now, we recall that $\ct=\ce$, because these are the 
velocities of the light defined with respect to the corresponding
sources at rest. 

We also recall that special relativity  
is based on the assumption of the
conservation of the rectilinear uniform movement.
But it is important to notice that, in general, reception of light does not 
imply in the conservation of the relative velocity, between 
$\Oe$ and $\Ot$, given by $\vt=\ve$. So, here, we first assume that
\BEQ \label{18} \Mwt(\tt)= \Mw(\te). \EEQ
This expresses the conservation of the number of cycles counted at $\Ot$,
for both observers.
Then, using equations (15) and (16), we find 
\BEQ \label{19} \betat=\frac{\vt}{\ct}=\frac{\ve}{\ce}=\betae.  \EEQ

We can substitute these results in equation (14), to obtain
\BEQ \label {20} \Mwt(\tt)=(1-\betae)\Nw(\te). \EEQ

This equation, which relates explicitly the two observers, is
the fundamental form for the generalization of 
Galilean equivalence, for different observers. It relates the  
number of wave cycles which vary in time, between moving objects, 
for different observers,
in an unique observational experience. 

We can divide both sides of equation (20) by $\te$ finding:
\BEQ \label {21} \frac{\Mwt(\tt)}{\tt}=\gamma(1-\betae)\frac{\Nw(\te)}{\te}. \EEQ
Now we have a relation between the two separated frequencies in space, not depending on time.
\BEQ \label{22} \fwott= \gamma (1-\betae)\fwoee, \EEQ 
where the left side of this equation is the frequency of cycles of $\Oe$, 
for $\obst$. And now we 
can discuss the other relation 
of $\obst$ with the original source $\Oe$.

\subsection{ Postulate II: Conservation of information.}

The second postulate is about the classical observer as a limit
case of receiver. In classical physics, there is no
loss of information concerning material processes.

In our situation, the formation of the cycles at $\Oe$, originating the waves out of 
it, are observed according to the sequence $\Nc(\te)=n \te/sec$, for $n$ a certain
fixed number. This expression must be conserved, with respect to both observers, 
in their respective observational times. This condition is named conservation
of information about material facts. As usual, information is defined as a numerical function 
of the object, as introduced by Claude Shannon, in the physical literature.
\footnote{ In {\it A mathematical Theory of Communication} ~\cite{sh1948}, Shannon discusses the situation
of two physical systems coupled by a pattern of waves, and what
he assumes to be information. There he says that information is
the number of messages, or any monotonic function of this number. By message, 
he means a physical specification. Applying to our situation, 
the event is the formation of cycles and this process is described by
a numerical function, as he requires. }.
Now the object `material system', detached from its trajectory, 
is only defined as number of cycle. So, this is just an application or use of the 
usual concept of information in the conditions of the phenomenon we 
discuss~\cite{poni2006}.

Now we may consider that, in the point of view of
$\obst$, a register of waves at
$\Ot$ can also be considered as a source at rest.
The embankment has no limit in length and this permits to consider a
point $\Oe'$ approaching $\Ot$ with velocity $\ve$,
(in both points of view), according to the arguments of the previous section
V-B. Then, at this point on the embankment, it is possible to define a quantity given by
\BEQ \label {23} \Mwolet(\tt)=(1+\betae)\Mwt(\tt). \EEQ
As in equation (15), there is no difference in time since we are in the context
of $\obst$.

However, in the relativistic context, we can relate the two observers, respecting the
observational quantities. We can substitute the expression of $\Mwt(\tt)$, given by equation (20), yielding
\BEQ \label {24} \Mwolet(\tt)=(1+\betae)(1-\betae)\Nw(\te). \EEQ
Knowing that all these numerical  functions are linear functions
of time, we have 
\BEQ \label {25}  \Nw(\te)=\gamma\Nw(\tt), \EEQ 
and substituting this result in equation (24), we find

\BEQ \label {26} \Mwolet(\tt)=\gamma(1+\betae)(1-\betae)\Nw(\tt). \EEQ

In the relativistic point of view, the same delay in time of the 
previous case, for the separation between $\Oe$ and $\Ot$ must be assumed,
since for what matters there is always a delay in time for the receiver,
when it means to observe what goes on outside the train. Independently of
being approaching or departing, this delay only depends on the velocity. 

Then, using the principle of conservation of information,
we must have
\BEQ \label {27}  \Mwolet(\tt)=\Nw(\frac{\tt}{\gamma}) \EEQ
As it was said before, the form of the equation is conserved, but the
two observers are separated in time\footnote{This condition is not valid for material waves, only for light.}.

Using the fact that  
\BEQ \label {28}  \Nw(t)=\Nc(t), \EEQ 
for any time, we can substitute these two last equations in
equation (26), to arrive at

\BEQ \label{29}\gamma^2= \frac{1}{(1+\betae)(1-\betae)}. \EEQ

Substituting this expression in equation (17), we have the
so called Lorentz transformation for times, in its fundamental form:

\BEQ \label{30}\te= \frac{\tt}{(1-\frac{\ve^2}{\ce^2})^\frac{1}{2}}. \EEQ

\subsection{The Constancy of the Velocity of Light}

Now, that we found the first Lorentz transformation, independently
of the constancy of the velocity of the light, we can actually derive the constancy
by calculating the velocities of approaching and leaving the train 
in the relativistic point of view.

As we assumed by principle, $\obst$ is a classical observer and by this reason
he cannot measure relativistic effects such as constancy of
the velocity of the light. This constancy is only found when one
calculates the velocities of the light approaching
and leaving $\Ot$ in a mixed point of view. To complete the previous list
of concepts of velocity of light, we must include the 
following concepts:

$\coeotr$ is the velocity of the
light approaching $\Ot$ from $\Oe$ with respect to the reference $\tt$.

$\coeotr$ is the velocity of the 
light reaching $\Oe'$ from $\Ot$ with respect to the reference $\te$.

We may start from the relation 
$\ce=\ct$, where the right side is the velocity of the wave
emitted from $\Ot$, with respect to $\Ot$, in both points of view, 
including the classical view of $\Obst$.

Consequently, we can assume that there is
a $\lambda_t$ such that  
\BEQ \label{31} \frac {\lambda_e}{\lambda_t}=\frac{\te}{\tt} , \EEQ
holds. And substituting equation (30) in the last equation, we also find
a relativistic correction between the wave lengths of these waves:

\BEQ \label{32} \lambda_e=\gamma\lambda_t \EEQ

However, as said before, the velocity of the light approaching $\Ot$ must be 
calculated from a mixed point of view and a different correction holds for the approaching
wave length. 

Until now, only number of cycles were under the rules of relativity,
according to the two relativistic principles. But if we assume that a wave is placed between 
$\Oe$ and $\Ot$, conditions must also hold for the wave lengths in between the same origins.
Since there is an intrinsic time displacement due to the observational difference, thee must also be
an intrinsic effect on the wavelength due to the displacement of the source.
This corresponds to an intrinsic dilatation of the original wave-length. It has the
contrary effect in this case, because it is the source $\Oe$ which is relativistic
displaced with respect to the reference $\Ot$. 

Substituting these results, the first velocity is given by 
\BEQ \label{33} \coeotr=\frac{ \Mw(\te)(1+\betae)\lambda_e\gamma}{\tt}=\ce \EEQ
where the coefficient $(1+\betae)$ of $\lambda_e$ is the classical increase on the
wave length because of the movement of the source away from $\Ot$ and the factor  
$\gamma$ is the relativistic effect of length dilatation on the approaching wavelength. 

And the second velocity only includes a relativistic correction for the wave length, 
of the other wave, the one leaving $\Ot$, 
since the source is at rest. It is given by
\BEQ \label{34} \coeotr=\frac{\Nctt(\te)(\lambda_e/\gamma)}{\te}=\ce. \EEQ

We finally derived the constancy of the velocity of the
light, showing that it is a specific relativistic result only, not shared by any other
point of view.

\section{Conclusions}

In our set-up we discriminate three classes of observers: 
observers of light represented by the equations of $\obse$; receivers of light represented 
by the equations of $\obst$; and relativistic observers, represented by
the unifying equations.
The constancy of the velocity of the light is a relativistic result only. 
And the only one who observes the relativistic phenomena is the relativistic observer. 
But if the classical and the semi-classical observers do not exist, the relativistic
observer disappears too. And the relativistic world vanishes with this being. 

The train/embankment experiment is a prototype of relativistic phenomena not including interaction. 
The basic set up of the experiment
of Michelson and Morley is the same situation. We can consider the same train carrying two mirrors, 
at equal distances from the origin 
$\Ot$  of our train. One mirror is localized on a line perpendicular to the velocity of the train while
 the other is in the direction of the movement. In this situation, light from the origin $O'$  
of the train is sent back and forth a great number of times from this origin, which moves with velocity $v$. 
According to the experiment, there is no difference in phase between the two perpendicular beams when they 
meet back in the origin. 

In this experiment, the Earth corresponds to the train. Its tangential velocity corresponds to our $v$.
 The complete experiment is done under our single conventional Earth time, and we are the classical observers 
inside the train. Consequently, what the experiment confirms is that under its conditions, the classical limit
holds. And, what is the classical limit? It is just the condition in which discrepancies in the velocity of light 
are not shown. Then, the first step to measure a discrepancy in the velocity of the light is by not taking its
constancy for granted. But, in this case, we must find new ways to recognize, at remote distances, the objects 
of observation, because the constancy is just a theoretical tool for this.  
 
In this paper we showed that because classical observers are receivers of light the velocity of light {\it in vacuum} 
is absolute. We did not show that classical observers exist as absolute beings.

\section{ Acknowledgments}
The authors thank Igor Volovich for stimulating discussions, Hans Groeneveld for suggestions, 
Ernst Binz for drawing our attention to reference~\cite{si2007} and Peter Keefe for a careful
reading of the text. C. Pombo also thanks Jose Helayel-Neto, Jose Thadeu Cavalcante, Eduardo Paquet, Carlos Baladron Garcia,
David P. Costa and Aad van den Enden for inspiring discussions on the relation between physics and psychology.

\section*{Appendix: A model for physical observers} 

The purpose of this appendix is to discuss a model for physical observers borrowing elements from
analytical psychology and philosophy. In doing so, we shall partly follow steps of physicists like
Wolfgang Pauli, who introduced this psychology in the interpretation of classical physics, and mathematicians 
like Hermann Weyl, who concluded that the ultimate essence of the concept of the
Ego is the system of reference.

Physicists usually consider observation identical to acquisition of data. This opinion has 
been rather disorienting, considering the level of antagonism still present in the literature in discussions 
about relativistic and quantum realities. Collection of data is not what observation is, although sets of 
discrete or continuous data can be considered elements of physical observations.
A physical observation is an expression of a language made of observational concepts.
\footnote{These observational concepts need not be physical quantities, but we will restrict ourselves to 
the case where they are.}  
This language is made by sets of sentences such as equations, specifying the way these concepts are organized 
and related. In the simplest case, when the concepts are physical quantities, the relations are just associations 
between sets of data. The simplest examples of observation are equations of trajectories of moving particles.

But with sets of positions and measures of time alone one cannot arrive at these equations, because a trajectory
 involves a third element which is the movement of a body. It is the movement that settles the association and 
these trajectories are observed independently of the existence of any theory. Newtonian mechanics, which is a 
complete theory and does not only consist of observational expressions, was the first theory made for explaining 
these trajectories. Soon thereafter, new concepts emerged to describe global aspects of classical behavior of many bodies,
 in which the concept of trajectory does not play the main role. Observation of continuous distributions of matter 
such as fluids and waves produced other classical theories. From this point, physics developed in the direction of
 much more complex organization of data. Therefore it is very difficult to avoid the conclusion that physical 
observations are human acquisitions of language, where these languages are not the conventional ones but expressions
 of a collectivity, in the sense which we explain below.

To observe is also to think or have thoughts which can be linguistically expressed. 
Thoughts are complex formations of psychological elements existing and developing 
in intervals of time. This gives another meaning for a sentence of language, not as a fixed object made of signals, 
but as a constraint between elements during an interval of time. 
Consciousness has deep relations with language and therefore observers can also be considered from a psychological
point of view, as conscious experiences, belonging to specific classes of conscious phenomena. Not only, physical
observations must be understood as a special case of human expressions, in which all the subjective aspects of our
existence are eliminated. 

For these reasons, we take the term `psychological' strictly from the Analytical Psychology, developed by
Carl Gustav Jung and many collaborators, between them W. Pauli. 
This is the only psychological theory involved with a collective basis of the human thought~\cite{ju1959}.  
In this theory, consciousness is a process made of functions such as thinking, intuition, 
feeling, sensation and perception. The totality of the conscious experience can have elements from 
many different functional sources, that means,
descriptions involving sensations with feelings, between others. 

Thinking is one of the functions of consciousness, the one which associates `names' and `meanings' to experiences.
According to Jung there are two kinds of thinking: fantasy and language. This second kind of thinking can also 
include descriptions of subjective experiences involving sensation and emotions which are also experiences related
to the body of the observer. But it can also be involved only with objective elements. The function named 
perception would be thinking of sensations and it can also include physical concepts. However, these experiences
are always grounded on the spatial point of views of observers, what makes these reports subjective. 
This is not a feature of physical observations which are grounded on physical (objective) systems of reference. 
In our model we focus on `physical thoughts', eliminating all the other functions, to relate to the equations 
representing the physical observations.

The main feature of analytical psychology, which is also a {\it sine qua non} condition for any model of observer in 
physics, is the fact that it assumes an archetypal or collective base for the consciousness and consequently for
 thinking. In the context of this psychology, certain manifestations of consciousness are not acquisitions of 
individual humans but of the psychic history of human kind. This makes a great difference between analytical 
psychology and all other psychologies, including psychoanalysis of Sigmund Freud. It postulates the existence 
of an autonomic psyche, without any influence from the individual consciousness, from which the conscious experience
 takes its deepest meanings. This psychic level does not consist of formed experiences  but only of fundamental 
meanings or `frames' for organizing them, the so called `archetypal patterns'. And this permits the definition of
 objective experiences in a collective sense, only made by special concepts directly related to these patterns. 
This objectivity based on collective agreement is the main feature of physical observations. Actually, physical
 experiences are the most clear manifestations of a collective psyche. 

Pauli~\cite{paju1995}
wrote about the influence of archetypal patterns on the ideas of Johannes Kepler and on the foundations of
 classical physics. And a deep connection between physical and psychological experiences was made by  
Weyl, who recognized the psychic nature of the physical system of reference ~\cite{we1949}. 
He just identified our `physical frames of reference' with `components of the ego'. And the center of the consciousness,
 in analytical psychology, is the collective essence of the Ego, made mainly by the archetypal patterns 
(which are the psychological correlates of the innate ideas of Immanuel Kant ~\cite{ka1929} and of 
the intuitions of Luitzen Brouwer~\cite{br1907}). 
Considering only the patterns involved with physical experiences, we may use the expression `physical ego' only for a 
sector of the ego as defined by Jung. This definition of physical ego, generating collective experiences, 
gives rise to experiences which could 
be interpreted as belonging to a collective consciousness. The existence of the latter can be considered as paradoxical, 
and was firstly rejected by Jung. But during his life, Jung refined several times the concept of ego~\cite{ev2004},
reaching this final conception in which the ego is a collective and structural base for experiences.

The concept of physical ego, as adopted here, means a class of physical references,
as building blocks of physical experiences and consequently of the physical world. 
Each class may consist of infinite systems of the same references. And different classes 
consist of different sets of references. 
For Weyl, there could be no objective experience detached from the ego. 
``The objectization, by elimination of the ego ... 
does not fully succeed, and the coordinate system remains ..." ~\cite{we1949} 
We can assume here that system of coordinate is used meaning system of reference. This idea of Weyl 
makes it also possible to differentiate 
between classes of physical observers based on different classes of references. 
A class of references forms a closed structure 
in itself, classifying physical experiences and observational language as well. Here the 
term `world' accounts for the infinite set of all possible 
descriptions by a class of observers. 

The idea of Weyl, also permits to understand 
the relation between the different classes of observers from a different perspective. The 
concept of complex, introduced by 
Jung and collaborators, 
is the key for this analysis. A complex can be defined as a set of experiences limited by a specific ego. 
Comparing to our case, a complex would be represented by a set of observations corresponding, not
to different observers from the same class, but to a specific class of observers.
In Newtonian physics, for example, these references only consist of the set of real numbers, three spacial 
axis, time and movement. With these fundamental ingredients, the classical world is formed. This
would be the classical complex. With special relativity 
a new reference emerges, that is light, resulting in a deep change of reality. These new experiences,
belonging to a new integrative complex, consist of the observations of 
another kind of observer, named the relativistic observer, see Section V. 

The main issue of the present paper has been to point at different kinds of observers, 
coexisting inside the final experience.
The situation actually resembles human psychology and integration of personalities. 
It is well known that human personalities can be broken in psychological complexes.
Complexes have great importance in the understanding of consciousness. As practically independent 
psyches, their split accounts 
for situations such as double consciousness and schizophrenia. And, in principle, conditions like this can be 
cured, by means of a reintegration of the parts. It can also happen by means of the emergence of a third 
entity, giving an unity for them. This would be considered as a cure of
the pathological situation.

Here, the situation resembles these psychological
situations, but cannot be considered the same in the sense that we are not discussing processes 
of individual minds.
These complexes would not be human personalities but some sort of `collective manifestation'. 
Moreover, it is not
under a unique time that happens their integration. 
Our third observer, is not to be thought to
represent experiences based on a third reference of time or give
a special status for one of the references of time. It keeps two times as an integral feature
of the experience.

A previous inventory of the plural observation inside special relativity, 
was carried out by Bruno Latour ~\cite{la1988}. It is very interesting to realize that 
the original duplicity of languages of special relativity with the realization of an integration,  
needing a `personification', is clear from very different perspectives. In both approaches,
the existence of special relativistic events, objective or not, only can be
meaningful with respect to a third observer. More recently, an analysis of observational 
discrepancies in Einstein's exposition
of the train/embankment experiment, was made by Avi Nelson~\cite{ne2003}, and discussed by
other authors ~\cite{romane2004}. Nelson also considered the possibility of an observational plurality 
but kept the notion of observation attached to 
sensation and perception, in the way Einstein did. An integration between observers
leading to a third observer is not possible in these terms.

This is mainly because sensation and consequently perception are not processes of a 
collectivity but of individual organisms. 
Physical observations have an unique status and
cannot even be assumed to be neuro psychological, if we had to consider each of us individually as observers. 
Since the brain, considered as  
the set of neuro activity, is a physical object in space and time, it becomes very difficult,
and even a nonsense, to trace processes inside it, which could account for the objectivity of
the space outside, as an individual experience. 
But, considering the geometrical aspect of the experience of space only, another
context for the mind-body question emerges. 
Models for neural correlates of consciousness (NCC) involve processes of neural integration, 
considering single brains ~\cite{kocr2003,ch2000,si2007,coth..}. 
There is no successful model 
till now although evidences of neural processes mediating psychophysical processes have 
been found. If the neurological brains 
are just big complexes of interconnected neural systems, it is natural to think that integrations 
could also happen interconnecting different brains. As far as we know, this kind of parallelism 
has not yet been discussed as a possibility. But this type of 
speculation is out of the scope of the present paper.

Another crucial contribution 
for this model of observer adopted here, comes from the philosopher Rudolf Carnap~\cite{ca1964}, who stressed the linguistic 
nature of physical observations. For him, physical observations form `worlds of languages' by producing 
specific languages. These so called observational languages ought to be detached from the theoretical languages, 
inside specific theories. But the propose of Carnap apparently failed in practical terms because of the difficulty
 to find general criteria for separating observational concepts from the theoretical ones. He was the first to 
recognize this fact, pointing to the divergences of opinion on this matter. 

For finding the observational languages we need two criteria.  The first is to separate observational terms inside
a specific theory differentiating observational and theoretical languages. We adopt a historical guidance and 
simply postulate the primary references from an analysis of Newtonian trajectories. Then we separate from the 
observational language all statements with concepts not built exclusively by these primary references. 
The second criterion is to define different theories by means of recognizing different observational languages. 
This can be done by assuming that a change in the number of primary references, not considering changes in the physical 
meaning of the references, means a different observational language. 

From an integration of these ideas we conclude that to observe is to describe, focusing our analysis on the equations. 
The description is a kind of organization of the reality. An observer is represented or manifested by the report, 
not by the body of a person. For this reason, observations do not depend on the position of persons (bodies of persons) 
neither observers have physical influence on the world. Another aspect of observers and of these archetypal 
references here discussed is that these references are not choices from human beings.
According to analytical psychology, the unconsciousness has its own dynamics, independent from the consciousness. 
The development of the ego, considering its collective essence, is not under the domain
of consciousness. In this specific aspect, the development of knowledge does not obey our will.  

The question of the nature of the psyche, whether it is a metaphysical entity and whether observers only can be human,
\footnote{\\If history had developed differently by not having the world dominated by us, homo sapiens-sapiens, 
it might have been necessary to question whether Neanderthals and Homo Erectus  could have served 
as good 'observers' in the sense discussed here. In reality, there remains this 
question about the more clear cut case chimpanzees and other great apes. Though the answer seems that they are not
`good observers' of the type we discussed, 
understanding of their language would be of great help to reach a definite conclusion.\label{footnoteApe}} 
deserves careful discussions but it can be avoided here, at the expense of understanding that language, as strongly 
emphasized by Karl 
Popper~\cite{po1974}, has an objective feature, which is independent of single human physical existences. Although being a
 product of human culture, it is made outside individuals. For this reason, in case of physical languages, observational 
information can be conveyed by instruments (but it is not to be misunderstood as being the instrument) independently of
 a physical presence of a specific person to read it. The only difference introduced here from analytical psychology 
is in the collective or archetypal nature of the observational concepts, with respect to the human observer, which 
contrasts with the opinion of Popper. This contrast is more related to the way in which we know and understand these
 concepts, whether from experimentation or not. But this is a very old philosophical discussion which we also do not
 need to enter. We just take for grant that these physical concepts of space and time have a collective meaning, 
that they do not belong to specific cultures but to the totality of the human kind. Lastly, we assume 
that for the situations discussed here, these observations consist of continuous associations between these concepts, 
what cannot be made by instruments or machines.

\end{document}